\documentclass[10pt,twocolumn,twoside]{IEEEtran}

\usepackage[final]{graphicx}
\usepackage[compress]{cite}
\usepackage[T1]{fontenc}
\usepackage{amsmath}
\usepackage{amsfonts}
\usepackage{amssymb, mathrsfs, amsthm}
\usepackage{latexsym}
\usepackage{algorithm}
\usepackage{algorithmic}
\usepackage{subfigure}

\IEEEoverridecommandlockouts

\begin{document}

\title{Convergence Constrained Multiuser Transmitter-Receiver Optimization in Single Carrier FDMA}

\author{Valtteri~Tervo*,~\IEEEmembership{Student Member,~IEEE,} Antti~T\"{o}lli,~\IEEEmembership{Member,~IEEE,} Juha~Karjalainen,~\IEEEmembership{Member,~IEEE,} \\ Tad~Matsumoto,~\IEEEmembership{Fellow,~IEEE} \\
{\bf EDICS}: MSP-MULT, MSP-CODR, OTH-NSP

\thanks{This work was supported by Finnish Funding Agency for Technology and Innovation (TEKES),
Academy of Finland, Riitta ja Jorma J.\ Takanen Foundation, Finnish Foundation for Technology Promotion, Walter Ahlström Foundation, Ulla Tuominen foundation and KAUTE-foundation. This work was also in part
supported by the Japanese government funding program, Grant-in-Aid
for Scientific Research (B), No. 23360170. A part of the material in this paper was presented at the Annual Conference on Signals, Systems, and Computers, Pacific Grove, CA, USA, November 4-7, 2012.

V.\ Tervo and T.\ Matsumoto are with the Centre for Wireless Communications, University of Oulu, P.O. Box 4500, 90014 University of Oulu, Finland, and Japan Advanced Institute of Science and Technology,
1-1 Asahi-Dai, Nomi, Ishikawa, 923-1292 Japan, email: \{wade, matumoto\}@ee.oulu.fi.

A.\ T\"{o}lli is with the Centre for Wireless Communications, University of Oulu, email: atolli@ee.oulu.fi.

J.\ Karjalainen is with the Samsung Electronics R\&D Institute UK, Falcon Business Park, Hali Building, Vaisalantie 4, 01230 Espoo, Finland, email: juha.pe.karjalainen@gmail.com.}}

\maketitle

%Page numbers
%\thispagestyle{plain}
%\pagestyle{plain}

\begin{abstract}

Convergence constrained power allocation (CCPA) in single carrier multiuser (MU) single-input multiple-output (SIMO) systems with turbo equalization is considered in this paper. In order to exploit full benefit of the iterative receiver, its convergence properties need to be considered also at the transmitter side. The proposed scheme can guarantee that the desired quality of service (QoS) is achieved after sufficient amount of iterations. We propose two different successive convex approximations for solving the non-convex power minimization problem subject to user specific QoS constraints. The results of extrinsic information transfer (EXIT) chart analysis demonstrate that the proposed CCPA scheme can achieve the design objective.  Numerical results show that the proposed schemes can achieve superior performance in terms of power consumption as compared to linear receivers with and without precoding as well as to the iterative receiver without precoding.

\end{abstract}


\section{Introduction}

Frequency domain equalization (FDE) for single-carrier transmission \cite{Falconer-Ariyavisitakul-Benyamin-Seeyar-Eldson-02} and multi-carrier schemes based on orthogonal division multiplexing (OFDM) \cite{Chang-66} are two efficient techniques for tackling the inter-symbol-interference (ISI) problem in frequency selective fading channels. Both of aforementioned techniques can be extended to multiuser communications yielding single-carrier frequency division multiple access (FDMA) \cite{Pancaldi-Vitetta-Kalbasi-Al-Dhahir-Uysal-Mheidat-08} and orthogonal frequency division multiple access (OFDMA) \cite{Nee-Prasad-00}, respectively. In OFDMA all available subcarriers are grouped into different subchannels\footnote{The bandwidth of each subchannel is less than the coherence bandwidth of the channel which results in flat fading subchannels.} that are assigned to distinct users. User separation at the receiver side is straightforward due to the orthogonality of the subchannels.

Single-carrier FDMA can be viewed as a form of OFDMA in which extra discrete fourier transform (DFT) and inverse DFT (IDFT) are added at the transmitter and receiver ends, respectively. DFT precoder spreads all the symbols across the whole frequency band forming a virtual single carrier structure. The advantage of FDMA as compared to OFDMA is its lower peak-to-average power ratio (PAPR). However, the optimal multi-user detection in single carrier FDMA in the presence of frequency selective channel results in prohibitive high computational complexity. A linear minimum mean squared error (LMMSE) detector provides an attractive low complexity scheme for the detection of FDMA signal in the presence of ISI and multiuser interference (MUI) utilizing the circulant structure of channel matrices \cite{Kansanen-Matsumoto-Schneider-Thoma-05, Yuan-Guo-Wang-Ping-08}.

Iterative FDE technique can achieve a significant performance gain over linear FDE in ISI channels \cite{Yuan-Guo-Wang-Ping-08}. In iterative FDE, the key idea is to utilize the feedback from a soft-output forward error correction (FEC) decoder that is updated according to "turbo" principle. To exploit the full merit of iterative receiver, the convergence properties of a receiver based on the "turbo" principle needs to be taken into account jointly at the transmitter and the receiver. In \cite{Shepherd-Shi-Reed-Schreckenbach-06}, extrinsic information transfer (EXIT) analysis \cite{tenBrink-01} is utilized to determine the optimal power allocation in a multiuser turbo coded code division multiple access (CDMA) system. In \cite{Yuan-Li-Ping-Lin-08}, the convergence analysis for MMSE based iterative equalizer is performed by using signal-to-noise (SNR) ratio variance charts \cite{Yuan-Guo-Wang-Ping-08}. Furthermore, the authors in \cite{Yuan-Li-Ping-Lin-08} use the convergence analysis to formulate the transmitter power allocation problem in frequency selective single-input single-output (SISO) channels with the iterative receiver mentioned above, assuming the availability of perfect channel state information (CSI) both at the transmitter and the receiver. In \cite{Karjalainen-Matsumoto-08, Karjalainen-Matsumoto-Utschick-08}, the impact of precoder design on the convergence properties of the soft cancellation (SC) frequency domain (FD) minimum mean-squared error (MMSE) equalizer is demonstrated. In \cite{Yuan-Xu-Ping-Lin-09}, precoder design for multiuser (MU) multiple-input multiple-output (MIMO) ISI channels based on iterative LMMSE detection is considered. The design criterion of the precoder in \cite{Yuan-Xu-Ping-Lin-09} is to maximize the signal-to-interference and noise ratio (SINR) at the end of the iterative process. In \cite{Karjalainen-Codreanu-Tolli-Juntti-Matsumoto-11}, in-depth analysis of the power allocation problem in single-carrier MIMO systems with iterative FD-SC-MMSE equalization has been presented.

EXIT chart is one of the most powerful tools for analyzing and optimizing parameters in iterative processing \cite{tenBrink-99, Ashikhmin-Kramer-tenBrink-04, Brannstrom-Rasmussen-Grant-05}. The convergence of an iterative process can be predicted by investigating the exchange of extrinsic information of the soft in / soft out (SftI/SftO) blocks in the form of mutual information (MI) of transmitted bits and the corresponding log-likelihood ratios (LLRs). This analysis can be made independently for each block which eliminates the necessity of time consuming chain simulations. When applied to joint equalizer and FEC decoder design, the objective is to guarantee an open convergence tunnel between the equalizer's and the decoder's EXIT function. To be more specific, the EXIT function of the equalizer has to be above the inverse EXIT function of the decoder until so called MI convergence point, which determines the communication reliability represented by bit error probability (BEP) achieved by the iterative equalizer. Therefore, the width of the tunnel as well as the MI convergence point are the key parameters when optimizing an iterative process using EXIT charts \cite{tenBrink-Kramer-Ashikhmin-04, tenBrink-Kramer-03}.

The contributions of this paper are summarized as follows: We extend the convergence constrained power minimization problem \cite{Karjalainen-Codreanu-Tolli-Juntti-Matsumoto-11} for multiuser (MU) single-input multiple-output (SIMO) system which results in joint optimization of multiple transmitters and the iterative receiver. The presence of multiple users makes the problem considerably more challenging due to the multidimensionality of the EXIT functions.
In \cite{Karjalainen-Codreanu-Tolli-Juntti-Matsumoto-11}, only quadrature phase sift keying (QPSK) modulation was considered. In this paper, we also derive a heuristic approach for 16-ary quadrature amplitude modulation (16QAM). The aim is to minimize the power consumption in single-carrier FDMA with iterative detection subject to quality of service (QoS) constraint. This can be adopted for example in long term evolution (LTE) type of systems \cite{3GPP-12}. Unlike in \cite{Karjalainen-Codreanu-Tolli-Juntti-Matsumoto-11} the joint optimization of the multiple transmitters and the receiver is not convex. Thus, we use block coordinate descent (BCD) method \cite{Bertsekas-99} where the non-convex joint optimization problem is split to separate transmitter and receiver optimization problems. Furthermore, we show that this type of alternating optimization converges to a local solution. Two efficient algorithms based on successive convex approximation (SCA) method \cite{Chiang-09} are proposed for solving the transmitter optimization problem for fixed receiver.

The rest of the paper is organized as follows: The system model of single carrier uplink transmission with multiple single-antenna users and a base station with multiple antennas is presented in Section \ref{sec: system model}. In Section \ref{sec: detector}, iterative frequency domain equalizer is described. Convergence constrained power allocation (CCPA) for turbo equalizer is derived in Section \ref{sec: CCP}. In Section \ref{sec: Algorithms}, the algorithms for solving the CCPA problem are derived.
The performance of proposed algorithms are demonstrated through simulations in Section \ref{sec: results}. Finally, conclusions are drawn in Section \ref{sec: conclusions}.

{\bf\textit{Nomenclature~}~--~}~Following notations are used throughout the paper: Vectors are denoted by lower boldface letters and matrices by uppercase boldface letters. The superscripts $^{\text{H}}$ and $^{\text{T}}$ denote Hermitian and transposition of a complex vector or matrix, respectively. $\mathbb{C}$, $\mathbb{R}$, $\mathbb{B}$ denote the complex, real and binary number fields, respectively. ${\bf I}_N$ denotes $N\times N$ identity matrix. The operator avg$\{\cdot\}$ calculates the arithmetic mean of its argument, diag$(\cdot)$ generates diagonal matrix of its arguments, $\otimes$ denotes the Kronecker product and $||\cdot||$ is the Euclidean norm of its complex argument vector.

\section{System Model} \label{sec: system model}

\begin{figure}[tbp!]
\centering
\includegraphics[angle=-90, width=\columnwidth]{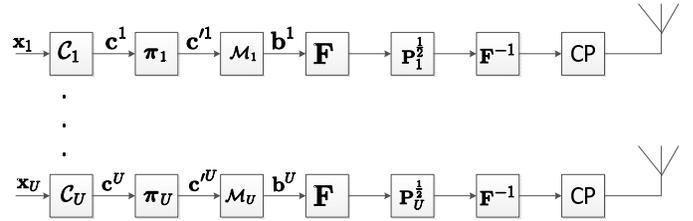}
\caption{The block diagram of the transmitter side of the system model.}
\label{fig: system model_TX}
\end{figure}

Consider uplink transmission with $U$ single antenna users and a base station with $N_R$ antennas. The transmitter side of the system model is depicted in Fig.\ \ref{fig: system model_TX}. Each user's data stream $\mathbf{x}_u\in\mathbb{B}^{R_c^uN_QN_F}$, $u=1,2,\dots,U$, is encoded by FEC code  $\mathcal{C}_u$ with a code rate $R_c^u\le1$. $N_Q$ denotes the number of bits per modulation symbol and $N_F$ is the number of frequency bins in discrete Fourier transform (DFT). The encoded bits $\mathbf{c}^u=[c_1^u,c_2^u,\dots,c_{N_QN_F}]^{\text{T}}\in\mathbb{B}^{N_QN_F}$ are bit-interleaved by multiplying $\mathbf{c}^u$ by pseudo-random permutation matrix $\boldsymbol{\pi}_u\in\mathbb{B}^{N_QN_F\times N_QN_F}$ resulting a bit sequence $\mathbf{c}'^u=\boldsymbol{\pi}_u\mathbf{c}^u$. After the interleaving, the sequence $\mathbf{c}'^u$ is mapped with a mapping function $\mathcal{M}_u(\cdot)$ onto a $2^{N_Q}$-ary complex symbol $b_l^u\in\mathbb{C}$, $l=1,2,\dots,N_F$, resulting a complex data vector $\mathbf{b}^u=[b_1^u,b_2^u,\dots,b_{N_F}^u]^{\text{T}}\in\mathbb{C}^{N_F}$. After the modulation, each user's data stream is spread across the subchannels by multiplying $\mathbf{b}^u$ by a DFT matrix $\mathbf{F}\in\mathbb{C}^{N_F\times N_F}$, $\forall u=1,2,\dots,U$,\footnote{The same amount of frequency domain resources are assumed to be allocated for each user in a cell.} where the elements of $\mathbf{F}$ are given by
\begin{equation}
f_{m,l}=\frac{1}{\sqrt{N_F}}e^{(i2\pi (m-1)(l-1)/N_F)},
\end{equation}
$m,l=1,2,\dots,N_F$. Each user's data stream is multiplied with its associated power allocation matrix $\mathbf{P}_u^{\frac{1}{2}}$, where $\mathbf{P}_u=\text{diag}([P_{u,1},P_{u,2},\dots,P_{u,N_F}]^{\text{T}})\in\mathbb{R}^{N_F\times N_F}$, with $P_{u,l}$ being the power allocated to the $l$th frequency bin. Finally, before transmission, each user's data stream is transformed into the time domain by the inverse DFT (IDFT) matrix $\mathbf{F}^{-1}$ and a cyclic prefix is added to mitigate the inter block interference (IBI).

\begin{figure}[tbp!]
\centering
\includegraphics[angle=-90, width=\columnwidth]{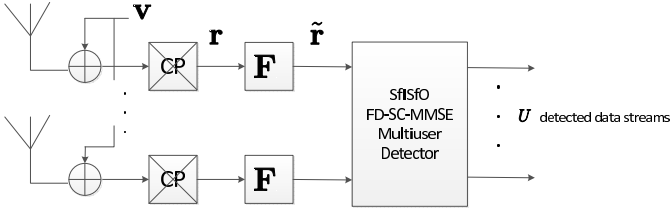}
\caption{The block diagram of the receiver side of the system model.}
\label{fig: system model_RX}
\end{figure}

The receiver side of the system model is depicted in Fig.\ \ref{fig: system model_RX}. After the cyclic prefix removal, the signal can be expressed as\footnote{In this paper, single cell scenario is considered and the impact of inter-cell-interference is excluded.}
\begin{equation} \label{eq: signal_after_CP_removal}
\mathbf{r}={\bf H}_u{\bf F}^{-1}\mathbf{P}_u^{\frac{1}{2}}\mathbf{F}\mathbf{b}^u+
\sum_{y\ne u}^{U}{\bf H}_y{\bf F}^{-1}\mathbf{P}_y^{\frac{1}{2}}\mathbf{F}\mathbf{b}^y+\mathbf{v},
\end{equation}
where
${\bf H}_u=[{\bf H}_u^1,{\bf H}_u^2,\dots,{\bf H}_u^{N_R}]^{\text{T}}\in\mathbb{C}^{N_RN_F\times N_F}$ is the space-time channel matrix for user $u$ and
$\mathbf{H}_u^r=\text{circ}\{[h_{u,1}^r,h_{u,2}^r,\dots,h_{u,N_L}^r,\mathbf{0}_{1\times N_F-N_L}]^{\text{T}}\}\in\mathbb{C}^{N_F\times N_F}$ is the time domain circulant channel matrix for user $u$ at the receive antenna $r$. The operator $\text{circ}\{\}$ generates matrix that has a circulant structure of its argument vector, $N_L$ denotes the length of the channel impulse response, $h_{u,l}^r$, $l=1,2,\dots,N_L$, $r=1,2,\dots,N_R$, is the fading factor of multipath channel.
%$\mathbf{F}_{N_R}=\mathbf{I}_{N_R}\otimes\mathbf{F}\in\mathbb{C}^{N_RN_F\times N_RN_F}$ is the unitary block DFT matrix.
A vector $\mathbf{v}\in\mathbb{C}^{N_RN_F}$ in \eqref{eq: signal_after_CP_removal} denotes white additive independent identically distributed (i.i.d.) Gaussian noise vector with variance $\sigma^2$. The signal $\mathbf{r}$ is transformed into the frequency domain by using DFT matrix $\mathbf{F}_{N_R}=\mathbf{I}_{N_R}\otimes\mathbf{F}\in\mathbb{C}^{N_RN_F\times N_RN_F}$, resulting
\begin{equation}
\tilde{\mathbf{r}}=\boldsymbol{\Gamma}\mathbf{P}^{\frac{1}{2}}\mathbf{F}_{U}\mathbf{b}+\mathbf{F}_{N_R}{\bf v},
\end{equation}
where $\boldsymbol{\Gamma}=[\boldsymbol{\Gamma}_1,\boldsymbol{\Gamma}_2,\dots,\boldsymbol{\Gamma}_{U}]
\in\mathbb{C}^{N_RN_F\times UN_F}$ and $\boldsymbol{\Gamma}_u=\text{bdiag}\{\boldsymbol{\Gamma}_{u,1},\boldsymbol{\Gamma}_{u,2},\dots,
\boldsymbol{\Gamma}_{u,N_F}\}\in\mathbb{C}^{N_RN_F\times N_F}$ is the space-frequency channel matrix for user $u$ expressed as
\begin{equation}
\boldsymbol{\Gamma}_u=\mathbf{F}_{N_R}{\bf H}_u\mathbf{F}^{-1},
\end{equation}
and $\boldsymbol{\Gamma}_{u,m}\in\mathbb{C}^{N_R\times N_R}$ is the diagonal channel matrix for $m^{\text{th}}$ frequency bin of $u^{\text{th}}$ user.
The power allocation matrix is composed by $\mathbf{P}=\text{diag}(\mathbf{P}_1,\mathbf{P}_2,\dots,\mathbf{P}_{U})\in\mathbb{R}^{UN_F\times UN_F}$, $\mathbf{F}_{U}=\mathbf{I}_{U}\otimes\mathbf{F}\in\mathbb{C}^{UN_F\times UN_F}$, and $\mathbf{b}=[{\mathbf{b}^1}^{\text{T}},{\mathbf{b}^2}^{\text{T}},\dots,{\mathbf{b}^{U}}^{\text{T}}]^{\text{T}}
\in\mathbb{C}^{UN_F}$.

\section{Receiver} \label{sec: detector}

%\subsection{FD-SC-MMSE turbo equalization}

\begin{figure}[tbp!]
\centering
\includegraphics[angle=-90, width=\columnwidth]{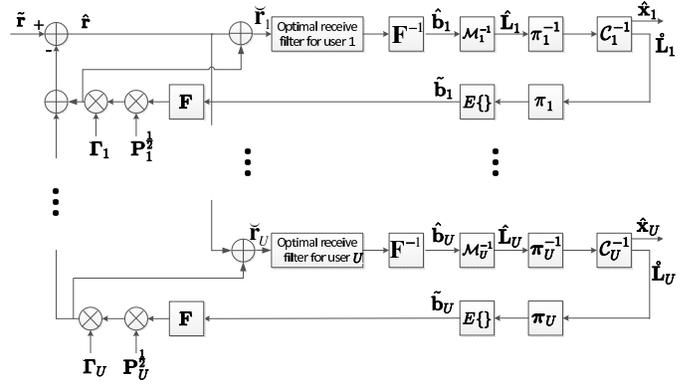}
\caption{The block diagram of frequency domain turbo equalizer.}
\label{fig: FD-SC-MMSE}
\end{figure}

The block diagram of the frequency domain turbo equalizer is depicted in Fig.\ \ref{fig: FD-SC-MMSE}. The frequency domain signal after the soft cancelation can be written as
\begin{equation}
\hat{\mathbf{r}}=\tilde{\mathbf{r}}-\boldsymbol{\Gamma}\mathbf{P}^{\frac{1}{2}}\mathbf{F}_{U}\tilde{\mathbf{b}},
\end{equation}
where $\tilde{\mathbf{b}}=[\tilde{\mathbf{b}^1}^{\text{T}},\tilde{\mathbf{b}^2}^{\text{T}},\dots,
\tilde{\mathbf{b}^{U}}^{\text{T}}]^{\text{T}}\in\mathbb{C}^{UN_F}$ is composed by $\tilde{\mathbf{b}}^u=[\tilde{b}_1^u,\tilde{b}_2^u,\dots,\tilde{b}_{N_F}^u]^{\text{T}}\in\mathbb{C}^{N_F}$. The soft symbol estimate $\tilde{b}_n^u$ is calculated as \cite{Karjalainen-Codreanu-Tolli-Juntti-Matsumoto-11}
\begin{equation}
\tilde{b}_n^u=E\{b_n^u\}=\sum_{b_i\in\mathfrak{B}}b_i\Pr(b_n^u=b_i),
\end{equation}
where $\mathfrak{B}$ is the modulation symbol alphabet, and the symbol \textit{a priori} probability can be calculated by \cite{Tuchler-Singer-Koetter-02}
\begin{align}
\Pr(b_n^u=b_i)&=\prod_{q=1}^{N_Q}\Pr({c'}_{n,q}^u=s_{i,q}) \nonumber \\
&=\Big(\frac{1}{2}\Big)^{N_Q}\prod_{q=1}^{N_Q}(1-\bar{s}_{i,q}\tanh(\lambda_{n,q}^u/2)),
\end{align}
with $\bar{s}_{i,q}=2s_{i,q}-1$ and ${\bf s}_i=[s_{i,1}, s_{i,2}, \dots, s_{i,N_Q}]^{\text{T}}$ is the binary representation of the symbol $b_i$, depending on the modulation mapping. $\lambda_{n,q}^u$ is the \textit{a priori} LLR of the bit ${c'}_{n,q}^u$, provided
by the decoder of user $u$.
After the soft cancelation, the residual and the estimated received signal of user $u$ are summed in  $\breve{\mathbf{r}}_{u}\in\mathbb{C}^{N_RN_F}$ as \cite{Karjalainen-11}
\begin{equation} \label{eq: rvankyra}
\breve{\mathbf{r}}_{u}=\hat{\mathbf{r}}+\boldsymbol{\Gamma}_u\mathbf{P}_u^{\frac{1}{2}}\mathbf{F}
\tilde{\mathbf{b}}^u.
\end{equation}
The time domain output of the receive filter for the $u$th user can be written as
\begin{equation} \label{eq: filter output}
\hat{\mathbf{b}}^u=\mathbf{F}^{-1}\breve{\boldsymbol{\Omega}}_u^{\text{H}}\breve{\mathbf{r}}_u,
\end{equation}
where $\breve{\boldsymbol{\Omega}}_u=[\breve{\boldsymbol{\Omega}}_u^1,\breve{\boldsymbol{\Omega}}_u^2,\dots,
\breve{\boldsymbol{\Omega}}_u^{N_R}]^{\text{T}}\in\mathbb{C}^{N_RN_F\times N_F}$ is the filtering matrix for the $u^{\text{th}}$ user and $\breve{\boldsymbol{\Omega}}_u^r\in\mathbb{C}^{N_F\times N_F}$ is the filtering matrix for $r^{\text{th}}$ receive antenna of $u^{\text{th}}$ user.
The effective SINR of the prior symbol estimates for $u^{\text{th}}$ user after FEC decoding can be expressed as
\begin{align} \label{eq: zeta}
\zeta_u=\frac{1}{N_F}\sum_{m=1}^{N_F}\frac{P_{u,m}{\boldsymbol{\omega}}_{u,m}^{\text{H}}{\boldsymbol\gamma}_{u,m}
{\boldsymbol\gamma}_{u,m}^{\text{H}}{\boldsymbol{\omega}}_{u,m}}{{\boldsymbol{\omega}}_{u,m}^{\text{H}}
{\bf\Sigma}_{\hat{\bf r},m}{\boldsymbol{\omega}}_{u,m}},
\end{align}
where ${\boldsymbol\gamma}_{u,m}\in\mathbb{C}^{N_R}$ consists of the diagonal elements of ${\boldsymbol\Gamma}_{u,m}$, i.e., ${\boldsymbol\gamma}_{u,m}$ is the channel vector for $m^{\text{th}}$ frequency bin of user $u$.  ${\boldsymbol{\omega}}_{u,m}=\Big[[\breve{\boldsymbol{\Omega}}_u^1]_{[m,m]},[\breve{\boldsymbol{\Omega}}_u^2]_{[m,m]}
\dots,[\breve{\boldsymbol{\Omega}}_u^{N_R}]_{[m,m]}\Big]^{\text{T}}\in\mathbb{C}^{N_R}$ is the receive beamforming vector for $m^{\text{th}}$ frequency bin of user $u$, and ${\bf\Sigma}_{\hat{\bf r},m}\in\mathbb{C}^{N_R\times N_R}$ is the interference covariance matrix of the $m^{\text{th}}$ frequency bin given by
\begin{equation} \label{eq: interference covariance}
{\bf\Sigma}_{\hat{\bf r},m}=\sum_{l=1}^{U}P_{l,m}
{\boldsymbol\gamma}_{l,m}
{\boldsymbol\gamma}_{l,m}^{\text{H}}\bar{\Delta}_l+\sigma^2{\bf I}_{N_R}.
\end{equation}
$\bar{\Delta}_l=\text{avg}\{\mathbf{1}_{N_F}-\ddot{\mathbf{b}}^l\}$ is the average residual interference of the soft symbol estimates and $\ddot{\mathbf{b}}^l=[|\tilde{b}_1^l|^2,|\tilde{b}_2^l|^2,\dots,|\tilde{b}_{N_F}^l|^2]^{\text{T}}\in\mathbb{C}^{N_F}$. %and $\bar{\ddot{\beta}}_u=\text{avg}\{\ddot{\bf b}^u\}$.

\section{Convergence Constrained Power Allocation} \label{sec: CCP}

In this section, the joint power allocation and receive beamforming optimization problem for iterative receiver is derived. The general problem formulation follows from \cite{Karjalainen-Codreanu-Tolli-Juntti-Matsumoto-11} where CCPA is derived for single user MIMO systems. However, the major difference compared to  \cite{Karjalainen-Codreanu-Tolli-Juntti-Matsumoto-11} is that the EXIT space now has $U+1$ dimensions which makes the problem considerably more challenging.

This section is outlined as follows: at first, the general problem formulation for multiuser SIMO systems is provided. We show that the convergence is guaranteed as long as there exist a tunnel between the $U+1$-dimensional EXIT surfaces.  After that, we introduce a novel \textit{diagonal sampling} approach which makes the problem solvable without performing exhaustive search. Then, we show how to transfer the MI constraints to LLR variance constraints in the case of BPSK and QPSK.  Finally, we apply CCPA to the case of 16QAM and show that the proposed convergence constraint guarantees the convergence also for 16QAM. Gray mapping is assumed throughout the derivation.

\subsection{General Problem Formulation} \label{Sec: General problem formulation}

\begin{figure}[tbp!]
\centering
\includegraphics[angle=-90,width=\columnwidth]{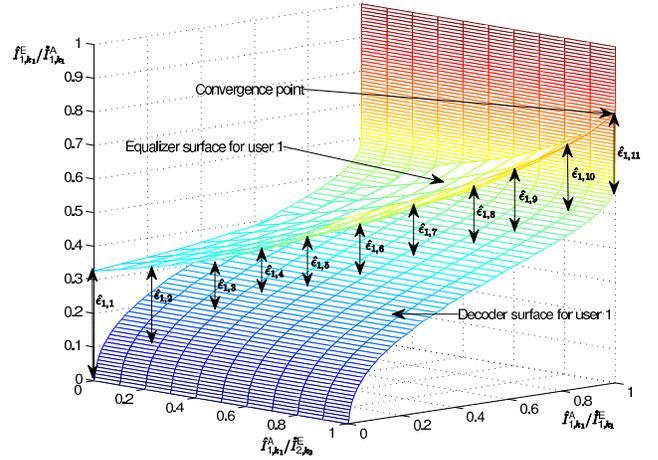}
\caption{An example of 3 dimensional formulation of the problem for user 1. $U=2$, $N_F=8$, $N_R=1$, $K=11$, $\hat{I}_u^{\text{E,target}}=0.8$, $\mathring{I}_u^{\text{E,target}}=0.9999$, $\epsilon_u=0.1$, $u=1,2$, $R_c=1/3$, $N_L=5$.}
\label{fig: 3D problem formulation}
\end{figure}

Let $\hat{I}_u^{\text{E}}$ denote the average MI between the transmitted interleaved coded bits ${\bf c}'^u$ and the LLRs at the output of the equalizer $\hat{\bf L}_u$ \cite[Eq. (18)]{Karjalainen-Codreanu-Tolli-Juntti-Matsumoto-11}. For notational convenience, equalizer refers to the combined block of the receive filter and soft mapper / demapper. Similarly to \cite{Karjalainen-Codreanu-Tolli-Juntti-Matsumoto-11} maximum a posteriori (MAP) soft demapper / mapper is used in this paper. Moreover, let $\hat{I}_u^{\text{A}}$ denote the \textit{a priori} MI at the input of the equalizer and $\hat{f}_u:[0,1]^U\rightarrow[0,1]$ denote a monotonically increasing EXIT function of the equalizer of the $u^{\text{th}}$ user.
Using similar definitions for the decoder of the $u^{\text{th}}$ user replacing $\hat{}$ with $\mathring{}$,
the essential condition for the convergence of the turbo equalizer can be written as
\begin{align} \label{eq: constr1}
\exists\{\mathring{I}_i^{\text{E}}\in[0,1]\}_{\substack{i=1 \\ i\ne u}}^U:
\hat{f}_u(\mathring{I}_1^{\text{E}},\dots,\mathring{I}_u^{\text{E}},\dots,\mathring{I}_{U}^{\text{E}})
\ge\mathring{f}_u^{-1}
(\mathring{I}_u^{\text{E}})+\epsilon_u \nonumber \\
\forall u=1,2\dots,U,
\end{align}
i.e., for all $u$, there exists a set of outputs from the decoders of all the users except $u$ such that the EXIT function of the equalizer of user $u$ is above the inverse of the EXIT function of the decoder of user $u$ plus a parameter $\epsilon_u$, which controls the minimum gap between the $U+1$-dimensional EXIT function of the equalizer of user $u$ and the inverse of the decoder's EXIT function of user $u$. In other words, the convergence is guaranteed as long as there exists an open tunnel between the two EXIT surfaces until the convergence point. The constraint \eqref{eq: constr1} is much more challenging to deal with than \cite{Karjalainen-Codreanu-Tolli-Juntti-Matsumoto-11} where the EXIT chart was 2-dimensional.
This is illustrated in the case of two users in Fig.\ \ref{fig: 3D problem formulation} where we can see the impact of the \textit{a priori} information coming from the other user's decoder.

We demonstrate that \eqref{eq: constr1} guarantees the convergence: Let $U=2$ and assume that there exists an open tunnel between the EXIT surfaces until the convergence point as presented in Fig.\ \ref{fig: 3D problem formulation}. Let $\mathring{I}_{u}^{\text{E,target}}$, $0\le\mathring{I}_{u}^{\text{E,target}}\le1$, be the target MI point of user $u$ after iterations. Furthermore, let $i_u\in\mathbb{N}$ be the index of iteration and $\mathring{I}_{u,i_u}^{\text{E}}$ denote the MI after iteration $i_u$ such that $\mathring{I}_{u,i_u+1}^{\text{E}}\ge\mathring{I}_{u,i_u}^{\text{E}}$.
Focusing on the user 1, the condition \eqref{eq: constr1} is written
\begin{align} \label{eq: MIconstrproof}
\hat{f}_{1}(\mathring{I}_{1}^{\text{E}},\mathring{I}_{2}^{\text{E}})
\ge\mathring{f}_1^{-1}
(\mathring{I}_{1}^{\text{E}})+\epsilon_{1},
\end{align}
such that for each $\mathring{I}_{1,i_1}^{\text{E}}$, $0\le\mathring{I}_{1,i_1}^{\text{E}}\le\mathring{I}_{1}^{\text{E,target}}$ there exists at least one $\mathring{I}_{2,i_2}^{\text{E}}$, $0\le\mathring{I}_{2,i_2}^{\text{E}}\le\mathring{I}_{2}^{\text{E,target}}$ that satisfies the condition.
Let the output value after the first activation of the decoder 1 be $\mathring{I}_{1,1}^{\text{E}}$, such that \eqref{eq: MIconstrproof} holds for some $\mathring{I}_{2,\tilde{i}_2}^{\text{E}}$. Due to the monotonicity of the EXIT function the condition \eqref{eq: MIconstrproof} holds for all indices $i_2\ge\tilde{i}_2$. Activating the decoder of user 1 again, the output of the equalizer becomes $\hat{f}_{1}(\mathring{I}_{1,2}^{\text{E}},\mathring{I}_{2,\tilde{i}_2}^{\text{E}})$. If the condition \eqref{eq: MIconstrproof} does not hold at the point $(\mathring{I}_{1}^{\text{E}},\mathring{I}_{2}^{\text{E}},\hat{I}_{1}^{\text{E}})=
(\mathring{I}_{1,2}^{\text{E}},\mathring{I}_{2,\tilde{i}_2}^{\text{E}},\hat{f}_{1}(\mathring{I}_{1,2}^{\text{E}},
\mathring{I}_{2,\tilde{i}_2}^{\text{E}}))$ in the 3-dimensional EXIT chart, i.e., $\hat{f}_{1}(\mathring{I}_{1,2}^{\text{E}},
\mathring{I}_{2,\tilde{i}_2}^{\text{E}}))<\mathring{f}_1^{-1}
(\mathring{I}_{1,2}^{\text{E}})+\epsilon_{1}$, there exists at least one $\mathring{I}_{2,i_2}^{\text{E}}$ that satisfies \eqref{eq: MIconstrproof}. Hence, $\tilde{i}_2$ can be increased, i.e., decoder 2 can be activated until the condition holds\footnote{If $U>2$, all the decoders (excluding the decoder of user 1) can be activated until \eqref{eq: MIconstrproof} holds.}. This can be repeated for all the points until the convergence point.

To make the problem tractable, continuous convergence condition \eqref{eq: constr1} is discretized and replaced with
\begin{align} \label{eq: MIconstr}
\exists\Big\{\mathring{I}_{i,k_i}^{\text{E}}\in[0,1]:k_i\in\{1,2,\dots,K\}\Big\}_{\substack{i=1 \\ i\ne u}}^U: \nonumber \\
\hat{f}_{u}(\mathring{I}_{1,k_1}^{\text{E}},\dots,\mathring{I}_{u,k_u}^{\text{E}},
\dots,\mathring{I}_{U,k_{U}}^{\text{E}})
\ge\mathring{f}_u^{-1}
(\mathring{I}_{u,k_u}^{\text{E}})+\epsilon_{u,k_u}, \nonumber \\
\forall k_u=1,2,\dots,K, \forall u=1,2\dots,U,
\end{align}
where $\epsilon_{u,k_u}=\epsilon_u$, $\forall k_u<K_u$ and $\epsilon_{u,K_u}=0$. Without loss of generality, we can assume $K_u=K$, $\forall u=1,2,\dots,U$, i.e., the number of discrete points in the EXIT chart is the same for all users. Furthermore, we will assume that $\mathring{I}_{u,k_u+1}^{\text{E}}>\mathring{I}_{u,k_u}^{\text{E}}$, $\forall k_u=1,2,\dots,K-1$, i.e., the indexing is ordered such that the MI increases with the index.

\subsection{Diagonal Sampling}

A 3-dimensional EXIT chart for user 1 is depicted in Fig.\ \ref{fig: 3D problem formulation} for the case of $U=2$. $\hat{I}_{1,k_1}^{\text{A}}/\mathring{I}_{u,k_u}^{\text{E}}$, $u=1,2$, denotes the \textit{a priori} information for the equalizer of the user 1 provided by the decoder of the user $u$. Double arrows with $\hat{\epsilon}_{1,k_1}$, $k_1=1,2,\dots,11$, are placed at the diagonal sample points where the condition \eqref{eq: MIconstr} is checked and $\hat{\epsilon}_{1,k_1}\ge\epsilon_{1,k_1}$. In this example, we have selected $K=11$ even though in many cases smaller $K$ is enough to guarantee the convergence. Intuitively, a sufficient value of $K$ depends on the shape of the decoder EXIT function. However, this is left as a future study.

The number of constraints in \eqref{eq: MIconstr} is $KU$. However, to find the optimal solution, we need to know how to pick up the optimal set of sample points from $\{\mathring{I}_i^{\text{E}}\in[0,1]\}_{\substack{i=1 \\ i\ne u}}^U$ for each $u=1,2,\dots,U$. For finding the best set of sample points, i.e., the path from origin to the convergence point which leads to a minimum power consumption, one should be able to check all the possible paths in $U+1$ dimensional EXIT space from origin to the convergence point and choose the one which gives the best result. This leads to a combinatorial optimization problem which is difficult to solve.

If the EXIT surfaces of the decoder and the equalizer do not intersect at any sampled point, the only active constraints are the ones where there is no \textit{a priori} information available from the other users. This can be justified by the fact that the EXIT function is monotonically increasing with its arguments, i.e., $\hat{f}_{u}(\mathring{I}_{1,k_1}^{\text{E}},\dots,\mathring{I}_{u,k_u}^{\text{E}},\dots,\mathring{I}_{U,k_{U}}^{\text{E}})
\le\hat{f}_{u}(\mathring{I}_{1,\tilde{k}_1}^{\text{E}},\dots,\mathring{I}_{u,\tilde{k}_u}^{\text{E}},
\dots,\mathring{I}_{U,\tilde{k}_{U}}^{\text{E}})$ if $\mathring{I}_{u,k_u}^{\text{E}}\le\mathring{I}_{u,\tilde{k}_u}^{\text{E}}$ $\forall u=1,2,\dots,U$.
In such a case, we can write the constraint \eqref{eq: MIconstr} as
\begin{align} \label{eq: worstcase}
\hat{f}_{u}(0,0,\dots,0,\mathring{I}_{u,k_u}^{\text{E}},0,\dots,0)
\ge\mathring{f}_u^{-1}
(\mathring{I}_{u,k_u}^{\text{E}})+\epsilon_{u,k_u}, \nonumber \\
\forall u=1,2\dots,U, \forall k_u=1,2,\dots,K.
\end{align}
This is the tightest possible constraint and it clearly cannot provide the best solution because with high probability there is another sampling which guarantees the convergence with lower power consumption. However, if the user does not know the modulation coding scheme (MCS), i.e., FEC code and modulation mapping, of other users at the transmitter, one may consider of using the constraint \eqref{eq: worstcase} to guarantee the reliable communication.

A pragmatic approach is to check only the points in the $U+1$-dimensional EXIT space where all the decoder's outputs are equal, i.e., we check the $K$ points on the line from the origin to the convergence point. Thus, we can write the constraint \eqref{eq: MIconstr} as
\begin{align} \label{eq: diag MIconstr}
\hat{f}_{u}(\mathring{I}_{1,k}^{\text{E}},\dots,\mathring{I}_{u,k}^{\text{E}},
\dots,\mathring{I}_{U,k}^{\text{E}})
\ge\mathring{f}_u^{-1}
(\mathring{I}_{u,k}^{\text{E}})+\epsilon_{u,k}, \nonumber \\
\forall k=1,2,\dots,K, \forall u=1,2\dots,U.
\end{align}
A sophisticated guess is that the active constraints lie on the line from the origin to the convergence point due to the smoothness of the decoder surface. We will denote this approach as \textit{diagonal sampling}.

\subsection{BPSK / QPSK}

Similarly to \cite{Karjalainen-Codreanu-Tolli-Juntti-Matsumoto-11}, the MI constraint of \eqref{eq: MIconstr} can be transformed to variance constraint using the approximation of the inverse of the so called J-function \cite{Brannstrom-Rasmussen-Grant-05}
\begin{equation} \label{eq: J_function}
\sigma_Z^2=J^{-1}(I_Z)\approx\Bigg(-\frac{1}{H_1}\log_2(1-I_Z^{\frac{1}{H_3}})\Bigg)^{\frac{1}{H_2}},
\end{equation}
where $\sigma_Z^2$ is the LLR variance, $I_Z$ is the MI and the parameters $H_1$, $H_2$ and $H_3$ can be found by least squares (LS) curve fitting with the constellation constrained capacity (CCC) equation \cite{Brannstrom-04}. Now, the MI constraint of \eqref{eq: diag MIconstr} can be written as
\begin{align} \label{eq: QPSKLLRvariance_constraint}
\hat{\sigma}^2_{u}(\mathring{I}_{1,k}^{\text{E}},\dots,\mathring{I}_{u,k}^{\text{E}},\dots,
\mathring{I}_{U,k}^{\text{E}})
\ge\mathring{\sigma}^2_{u,k},  \nonumber \\
\forall k=1,2,\dots,K, \forall u=1,2\dots,U,
\end{align}
where $\hat{\sigma}^2_{u}(\mathring{I}_{1,k}^{\text{E}},\dots,\mathring{I}_{u,k}^{\text{E}},\dots,
\mathring{I}_{U,k}^{\text{E}})=\text{J}^{-1}(\hat{f}_{u}(\mathring{I}_{1,k}^{\text{E}},\dots,
\mathring{I}_{u,k}^{\text{E}},\dots,
\mathring{I}_{U,k}^{\text{E}}))$, is the variance of the conditional LLR distribution at the output of the equalizer of user $u$ depending on the MI at the output of all the decoders and  $\mathring{\sigma}^2_{u,k}=\text{J}_2^{-1}(\mathring{f}_u^{-1}
(\mathring{I}_{u,k}^{\text{E}})+\epsilon_{u,k})$ is the variance of the conditional LLR distribution at the input of the decoder of user $u$ depending on the MI at the output of the decoder of user $u$.

In \cite{Karjalainen-Codreanu-Tolli-Juntti-Matsumoto-11}, a result presented in \cite{Ramon-Herzet-Vandendorpe-07} is used to find an analytical expression of the LLR variance at the output of the equalizer in the case of QPSK. We can use the same result by noting that $\bar{\Delta}_l$ in \eqref{eq: interference covariance} is a function of the output of the decoder of user $l$ and hence,
the SINR \eqref{eq: zeta} is a function of the outputs of the decoders of all the users $\zeta_{u}(\mathring{I}_{1,k}^{\text{E}},\dots,
\mathring{I}_{u,k}^{\text{E}},\dots,
\mathring{I}_{U,k}^{\text{E}})$. Equation \cite[Eq. (17)]{Karjalainen-Codreanu-Tolli-Juntti-Matsumoto-11} can be extended to the multiuser case as
\begin{align} \label{eq: varianceSINR}
\hat{\sigma}^2_{u}(\mathring{I}_{1,k}^{\text{E}},\dots,\mathring{I}_{u,k}^{\text{E}},\dots,
\mathring{I}_{U,k}^{\text{E}})= \nonumber \\
\frac{4\zeta_{u}
(\mathring{I}_{1,k}^{\text{E}},\dots,\mathring{I}_{u,k}^{\text{E}},\dots,
\mathring{I}_{U,k}^{\text{E}})}{1-\zeta_{u}(\mathring{I}_{1,k}^{\text{E}},\dots,
\mathring{I}_{u,k}^{\text{E}},\dots,
\mathring{I}_{U,k}^{\text{E}})\bar{\Delta}_{u,k}}.
\end{align}
Substituting \eqref{eq: varianceSINR} to \eqref{eq: QPSKLLRvariance_constraint} the convergence constraint is written as
\begin{align} \label{eq: QPSKconv_constraint}
\zeta_{u}(\mathring{I}_{1,k}^{\text{E}},\dots,\mathring{I}_{u,k}^{\text{E}},\dots,
\mathring{I}_{U,k}^{\text{E}})\ge\xi_{u,k}, \nonumber \\
\forall u=1,2\dots,U, \forall k=1,2,\dots,K,
\end{align}
where
\begin{equation} \label{eq: xi}
\xi_{u,k}=\frac{\mathring{\sigma}_{u,k}^2}{4+\mathring{\sigma}_{u,k}^2\bar{\Delta}_{u,k}},
\end{equation}
is a constant that depends on the FEC code.

\subsection{A Heuristic Approach for 16QAM}

Similarly to QPSK case, the MI at the output of the demapper can be transformed to the variance of the conditional LLR distribution by using \eqref{eq: J_function}. However, the parameters $H_1$, $H_2$ and $H_3$ are found by fitting the function \eqref{eq: J_function} with the corresponding 16QAM results \cite{Kansanen-Matsumoto-07}.
Let $\text{J}_2$ and $\text{J}_4$ denote the J-functions for QPSK and 16QAM, respectively.
With these notations, the MI constraint of \eqref{eq: diag MIconstr} in the case of 16QAM can be written as
\begin{align} \label{eq: LLRvariance_constraint_QAM}
\text{J}_4^{-1}(\hat{f}_{u}(\mathring{I}_{1,k}^{\text{E}},\dots,\mathring{I}_{u,k}^{\text{E}},
\dots,\mathring{I}_{U,k}^{\text{E}}))
\ge\text{J}_4^{-1}(\mathring{f}_u^{-1}
(\mathring{I}_{u,k}^{\text{E}})+\epsilon_{u,k}), \nonumber \\
\forall k=1,2,\dots,K, \forall u=1,2\dots,U.
\end{align}

The difference in the system model of different modulation schemes arises in the soft demapper. To achieve the final form of the convergence constraint in \eqref{eq: QPSKconv_constraint} we used the expression \eqref{eq: varianceSINR} where Gray mapped QPSK is assumed. In 16QAM this mapping between the SINR and the variance of the LLR distributions does not hold anymore. However, substituting the parameter values from \cite[Table I]{Kansanen-Matsumoto-07} to \eqref{eq: J_function}, it can be easily verified that $\text{J}^{-1}_4(I_Z)\ge\text{J}^{-1}_2(I_Z)$\footnote{Equality holds when $I_Z=0$ or $I_Z=1$.}, $\forall I_Z\in[0,1]$.
Using this result, we can obtain that when modulation order increases, larger LLR variance is needed to achieve the same SINR, i.e.,
\begin{align} \label{eq: upper_bound_proof}
\text{J}_{4}^{-1}(\hat{f}_{u}(\mathring{I}_{1,k_1}^{\text{E}},\dots,\mathring{I}_{u,k_u}^{\text{E}},\dots,
\mathring{I}_{U,k_U}^{\text{E}}))\ge \nonumber \\
\text{J}_{2}^{-1}(\hat{f}_{u}(\mathring{I}_{1,k_1}^{\text{E}},\dots,\mathring{I}_{u,k_u}^{\text{E}},\dots,
\mathring{I}_{U,k_U}^{\text{E}}))= \nonumber \\
\frac{4\zeta_{u}
(\mathring{I}_{1,k_1}^{\text{E}},\dots,\mathring{I}_{u,k_u}^{\text{E}},\dots,
\mathring{I}_{U,k_{U}}^{\text{E}})}{1-\zeta_{u}(\mathring{I}_{1,k_1}^{\text{E}},\dots,
\mathring{I}_{u,k_u}^{\text{E}},\dots,
\mathring{I}_{U,k_{U}}^{\text{E}})\bar{\Delta}_{u,k_u}}.
\end{align}
We can conclude that for 16QAM the convergence constraint \eqref{eq: QPSKconv_constraint} is conservative, i.e., the resulting EXIT curve of the equalizer is never above the true $\hat{I}_{u,k}^{\text{E}}$, $\forall u,k$. Hence, the convergence constraint \eqref{eq: QPSKconv_constraint} guarantees the convergence for 16QAM. It should be noticed that the difference in convergence constraint between the QPSK and 16QAM arises in \eqref{eq: xi} where $\mathring{\sigma}_{u,k}^2$ is obtained using either $\text{J}^{-1}_2$ or $\text{J}^{-1}_4$ depending on the modulation.

\section{Transmitter - Receiver Optimization} \label{sec: Algorithms}

In this section, algorithms for solving the transmitter-receiver (Tx-Rx) optimization problem is presented. In Section \ref{sec: cyclic descent}, the joint Tx-Rx optimization problem is split to separate transmitter and receiver optimization problems. The non-convex Tx optimization problem for fixed Rx is considered in Sections \ref{sec: SCAVC} and \ref{sec: SCAGP}.

The power minimization problem with the convergence constraint derived in the previous section is expressed as
\begin{equation}
\begin{array}{lll} \label{eq: SIMO_opt_original}
\underset{{\bf P}, \breve{\boldsymbol{\Omega}}^k}{\text{minimize}} & \text{tr}\{\mathbf{P}\} &  \\
\text{subject to}
& \zeta_{u}(\mathring{I}_{1,k}^{\text{E}},\dots,\mathring{I}_{u,k}^{\text{E}},\dots,
\mathring{I}_{U,k}^{\text{E}})\ge\xi_{u,k}, \\
&\forall u=1,2\dots,U, \forall k=1,2,\dots,K, \\
& P_{u,m}\ge0, \\
& u=1,2,\dots,U, m=1,2,\dots,N_F,
\end{array}
\end{equation}
where $\breve{\boldsymbol{\Omega}}^k$ is the receive filter at the $k^{\text{th}}$ MI index.

\subsection{Alternating Optimization} \label{sec: cyclic descent}

Our objective is to jointly optimize the power allocation at the transmitter and the beamforming vectors at the receiver while the convergence of the iterative receiver is guaranteed. Differentiating the Lagrangian of \eqref{eq: SIMO_opt_original} with respect to the receive beamforming vectors and equating to zero, the optimal receive beamforming vector for $m^{\text{th}}$ frequency bin of $u^{\text{th}}$ user at the $k^{\text{th}}$ MI index is given by
\begin{equation} \label{eq: opt receiver}
{{\boldsymbol{\omega}}_{u,m}^k}=\eta_{u}^k
\boldsymbol{\Sigma}_{\hat{\bf r},m,k}^{-1}\boldsymbol{\gamma}_{u,m}\sqrt{P_{u,m}},
\end{equation}
where $\eta_{u}^k\in\mathbb{R}$. Hence, the optimal receiver \eqref{eq: opt receiver} is actually the MMSE receiver used in \cite[Chapter 5]{Karjalainen-11} up to a scalar multiplier leading to exactly the same SINR. The scaling factor $\eta_{u}^k$ should be chosen such that it matches with the assumptions made in soft demapper.
With the notations given in Section \ref{sec: detector}, turbo equalizer works properly only if the scaling factor $\eta_{u}^k$ is chosen to be \cite{Kansanen-Matsumoto-07} $\eta_{u}^k=\frac{1}{\text{avg}\{\ddot{\bf b}^u\}\zeta_{u,k}+1}$.

The joint transmitter-receiver optimization problem can be solved by using the alternating optimization where we split the non-convex joint optimization problem to separate transmitter and receiver optimization. We start with a feasible initial guess\footnote{Can be found by e.g., using zero forcing algorithm \cite{Tse-Viswanath-05}.} $\hat{\bf P}^{(0)}$ and calculate the optimal receive filter. After that, the problem \eqref{eq: SIMO_opt_original} is solved for a fixed $\breve{\boldsymbol{\Omega}}^k$. A monotonic convergence of alternating optimization to a local optima can be justified by the fact that each step improves the objective. The overall algorithm is presented in \textbf{Algorithm} \ref{alg: block coordinate descent}, where ${\bf P}^{*}$ represents a solution of problem \eqref{eq: SIMO_opt_original} for fixed $\breve{\boldsymbol{\Omega}}^k$ and $\breve{\boldsymbol{\Omega}}^{k*}$ represents the optimal $\breve{\boldsymbol{\Omega}}^k$ for fixed ${\bf P}$. In the following sections, we will be focusing on solving the problem \eqref{eq: SIMO_opt_original} for fixed $\breve{\boldsymbol{\Omega}}^k$, denoted as power allocation problem (PAP).

\begin{algorithm}
\caption{Alternating Optimization.}
\label{alg: block coordinate descent}
\flushleft
\begin{minipage}{\textwidth}
\begin{enumerate}
\item
\begin{algorithmic}[1]
\STATE Initialize $\hat{\bf P}=\hat{\bf P}^{(0)}$
\REPEAT
\STATE Calculate the optimal $\breve{\boldsymbol{\Omega}}^k$ from \\ $\breve{\boldsymbol{\Omega}}^k_u=\frac{1}{\text{avg}\{\ddot{\bf b}^u\}\zeta_{u,k}+1}{\bf\Sigma}_{\hat{\bf r},k}^{-1}{\bf\Gamma}_u\hat{\bf P}_u^{\frac{1}{2}}$.
\STATE Set $\breve{\boldsymbol{\Omega}}^k=\breve{\boldsymbol{\Omega}}^{k*}$ and solve problem \eqref{eq: SIMO_opt_original} \\ with variables ${\bf P}$.
\STATE Update $\hat{\bf P}={\bf P}^{*}$
\UNTIL Convergence
\end{algorithmic}
\end{enumerate}
\end{minipage}
\end{algorithm}

To ease the handling of \eqref{eq: SIMO_opt_original}, we write the problem in equivalent form by splitting the convergence constraint as follows:
\begin{align} \label{eq: equiconstr}
&\frac{1}{N_F}\sum_{m=1}^{N_F}t_{u,m}^k\ge\xi_{u,k} \nonumber \\
&t_{u,n}^k=\frac{P_{u,n}|{\boldsymbol{\omega}_{u,n}^k}^{\text{H}}{\boldsymbol\gamma}_{u,n}|^2}
{
\sum_{l=1}^{U}P_{l,n}
|{\boldsymbol{\omega}_{u,n}^k}^{\text{H}}{\boldsymbol\gamma}_{l,n}|^2\bar{\Delta}_{k}+
\sigma^2||{\boldsymbol{\omega}_{u,n}^k}||^2}.
\end{align}
At the optimal point the constraints hold with equality and hence, we can relax the equality in \eqref{eq: equiconstr} leading to equivalent formulation
\begin{equation}
\begin{array}{lll} \label{eq: SIMO_opt_equivalent_orig}
\underset{{\bf P}, \breve{\boldsymbol{\Omega}}}{\text{minimize}} & \sum_{u=1}^{U}\sum_{m=1}^{N_F}P_{u,m} &  \\
\text{subject to} & \frac{1}{N_F}\sum_{m=1}^{N_F}t_{u,m}^k\ge\xi_{u,k} \\
& u=1,2,\dots,U, k=1,2,\dots,K,  \\
& \frac{P_{u,n}|{\boldsymbol{\omega}_{u,n}^k}^{\text{H}}{\boldsymbol\gamma}_{u,n}|^2}
{
\sum_{l=1}^{U}P_{l,n}
|{\boldsymbol{\omega}_{u,n}^k}^{\text{H}}{\boldsymbol\gamma}_{l,n}|^2\bar{\Delta}_{k}+
\sigma^2||{\boldsymbol{\omega}_{u,n}^k}||^2}\ge t_{u,n}^k, \\
& k=1,2,\dots,K, u=1,2,\dots,U, \\
& n=1,2,\dots,N_F, \\
& P_{u,n}\ge0, \\
& u=1,2,\dots,U, n=1,2,\dots,N_F.
\end{array}
\end{equation}

\subsection{Successive Convex Approximation via Variable Change} \label{sec: SCAVC}
Similarly to \cite{Kaleva-Tolli-Juntti-globecom12}, we introduce new variables $\alpha_{u,m}\in\mathbb{R}$, such that $P_{u,m}=e^{\alpha_{u,m}},\forall u=1,2,\dots,U,m=1,2,\dots,N_F$. The PAP with new variables can be equivalently written as
\begin{equation}
\begin{array}{lll} \label{eq: SIMO_opt_equivalent}
\underset{{\boldsymbol \alpha},{\bf t}}{\text{minimize}} & \sum_{u=1}^{U}\sum_{m=1}^{N_F}e^{\alpha_{u,m}} &  \\
\text{subject to} & \frac{1}{N_F}\sum_{m=1}^{N_F}t_{u,m}^k\ge\xi_{u,k} \\
& u=1,2,\dots,U, k=1,2,\dots,K,  \\
(**) & \frac{e^{\alpha_{u,n}}|{\boldsymbol{\omega}_{u,n}^k}^{\text{H}}{\boldsymbol\gamma}_{u,n}|^2}
{
\sum_{l=1}^{U}e^{\alpha_{l,n}}
|{\boldsymbol{\omega}_{u,n}^k}^{\text{H}}{\boldsymbol\gamma}_{l,n}|^2\bar{\Delta}_{k}+
\sigma^2||{\boldsymbol{\omega}_{u,n}^k}||^2}\ge t_{u,n}^k, \\
& k=1,2,\dots,K, u=1,2,\dots,U, \\
& n=1,2,\dots,N_F,
\end{array}
\end{equation}
where ${\bf t}=\{t_{u,m}^k: u=1,2,\dots,U,k=1,2,\dots,K,m=1,2,\dots,N_F\}$, and ${\boldsymbol \alpha}=\{\alpha_{u,m}: u=1,2,\dots,U,m=1,2,\dots,N_F\}$. Taking the natural logarithm of the constraint $(**)$ yields
\begin{align} \label{eq: logconstr}
&\alpha_{u,n}+2\ln(|{\boldsymbol{\omega}_{u,n}^k}^{\text{H}}{\boldsymbol\gamma}_{u,n}|) \nonumber \\
&-\ln(\sum_{l=1}^{U}e^{\alpha_{l,n}}
|{\boldsymbol{\omega}_{u,n}^k}^{\text{H}}{\boldsymbol\gamma}_{l,n}|^2\bar{\Delta}_{k}+
\sigma^2||{\boldsymbol{\omega}_{u,n}^k}||^2)\ge\ln t_{u,n}^k.
\end{align}
It is well known that logarithm of the summation of the exponentials is convex. Hence, the left hand side (LHS) of the constraint \eqref{eq: logconstr} is concave. The RHS of \eqref{eq: logconstr} can be locally approximated with its best convex upper bound, i.e., linear approximation of $\ln t_{u,n}^k$ at a point $\hat{t}_{u,n}^k$:
\begin{equation}
Y(t_{u,n}^k,\hat{t}_{u,n}^k)=\ln\hat{t}_{u,n}^k+\frac{(t_{u,n}^k-\hat{t}_{u,n}^k)}
{\hat{t}_{u,n}^k}.
\end{equation}
A local convex approximation of \eqref{eq: SIMO_opt_equivalent} can be written as
\begin{equation}
\begin{array}{lll} \label{eq: local_conv_appr}
\underset{{\boldsymbol \alpha}, {\bf t}}{\text{minimize}} & \sum_{u=1}^{U}\sum_{m=1}^{N_F}e^{\alpha_{u,m}}   \\
\text{subject to} & \sum_{m=1}^{N_F}t_{u,m}^k\ge N_F\xi_{u,k}, u=1,2,\dots,U, \\
& k=1,2,\dots,K, \\
& \alpha_{u,n}+2\ln(|{\boldsymbol{\omega}_{u,n}^k}^{\text{H}}{\boldsymbol\gamma}_{u,n}|)- \\ &\ln(\sum_{l=1}^{U}e^{\alpha_{l,n}}
|{\boldsymbol{\omega}_{u,n}^k}^{\text{H}}{\boldsymbol\gamma}_{l,n}|^2\bar{\Delta}_{k}+
\sigma^2||{\boldsymbol{\omega}_{u,n}^k}||^2)\ge \\ & Y(t_{u,n}^k,\hat{t}_{u,n}^k), u=1,2,\dots,U, \\
& k=1,2,\dots,K, n=1,2,\dots,N_F,
\end{array}
\end{equation}
and it can be solved efficiently by using standard optimization tools, e.g., interior-point methods \cite{Boyd-Vandenberghe-04}.

The SCA algorithm starts by a feasible initialization $\hat{t}_{u,n}^k=\hat{t}_{u,n}^{k(0)}, \forall u,k,n$. After this, \eqref{eq: local_conv_appr} is solved yielding a solution ${t}_{u,n}^{k(*)}$ which is used as a new point for the linear approximation. The procedure is repeated until convergence. The SCA algorithm is summarized in \textbf{Algorithm} \ref{alg: SCA}. By projecting the optimal solution from the approximated problem \eqref{eq: local_conv_appr} to the original concave function (RHS in \eqref{eq: logconstr}) the constraint becomes loose and thus, the objective can always be reduced. Hence, this algorithm is guaranteed to monotonically converge to a local optimum.

\begin{algorithm}
\caption{Successive convex approximation algorithm.}
\label{alg: SCA}
\begin{minipage}{\columnwidth}
\begin{algorithmic}[1]
\STATE Set $\hat{t}_{u,n}^k=\hat{t}_{u,n}^{k(0)}, \forall u,k,n$.
\REPEAT
\STATE Solve Eq.\ \eqref{eq: local_conv_appr}.
\STATE Update $\hat{t}_{u,n}^k={t}_{u,n}^{k(*)}, \forall u,k,n$.
\UNTIL Convergence.
\end{algorithmic}
\end{minipage}
\end{algorithm}

\subsection{Successive Convex Approximation via Geometric Programming} \label{sec: SCAGP}

Another algorithm for solving the PAP can be derived by using the approach introduced in \cite{Chiang-Tan-Palomar-Oneill-Julian-07} where the SCA is implemented via series of geometric programs (GPs) \cite{Boyd-Vandenberghe-04}. The inequality of weighted arithmetic mean and weighted geometric mean states that for any set of $\Phi_m,\alpha_m>0$, $m=1,2,\dots,N_F$,
\begin{equation}
\frac{\sum_{m=1}^{N_F}\Phi_m\alpha_m}{\Phi}\ge\sqrt[\Phi]{\prod_{m=1}^{N_F}\alpha_m^{\Phi_m}},
\end{equation}
where $\Phi=\sum_{m=1}^{N_F}\Phi_m$. Choosing $\Phi_m=\frac{\hat{t}_m}{\sum_{n=1}^{N_F}\hat{t}_n}$, $\hat{t}_m>0$, $m=1,2,\dots,N_F$, and denoting $\alpha_m=\frac{t_m}{\Phi_m}$, we have
\begin{equation} \label{eq: sumgemon}
\sum_{m=1}^{N_F}t_m\ge\prod_{m=1}^{N_F}(\frac{t_m}{\Phi_m})^{\Phi_m},
\end{equation}
for all $\Phi_m, t_m>0$, $m=1,2,\dots,N_F$. Therefore, the summation constraint can be replaced by its monomial underestimate and a local approximation of \eqref{eq: SIMO_opt_original} for fixed $\breve{\boldsymbol{\Omega}}^k$ can be written in the form of GP as
\begin{equation}
\begin{array}{lll} \label{eq: SISO_opt_3rd}
\underset{{\bf P}, {\bf t}}{\text{minimize}} & \text{tr}\{\bf P\}   \\
\text{subject to} & \prod_{n=1}^{N_F}(\frac{t_{u,n}^k}{\Phi_{u,n}^k})^{\Phi_{u,n}^k}\ge N_F\xi_{u,k}, \\
&u=1,2,\dots,U, k=1,2,\dots,K, \\
& P_{u,m}|{\boldsymbol{\omega}_{u,m}^k}^{\text{H}}\boldsymbol{\gamma}_{u,m}|^2\ge \\
& (\sum_{l=1}^{U}P_{l,m}|
{\boldsymbol{\omega}_{u,m}^k}^{\text{H}}\boldsymbol{\gamma}_{l,m}|^2
\bar{\Delta}_{k}+\sigma^2|{\boldsymbol{\omega}_{u,m}^k}|^2)t_{u,m}^k, \\
& u=1,2,\dots,U, k=1,2,\dots,K, \\
& m=1,2,\dots,N_F, \\
& P_{u,m}\ge0, \;\;\; u=1,2,\dots,U, m=1,2,\dots,N_F.
\end{array}
\end{equation}
Now the objective is a posynomial, the LHSs of the inequality constraints are monomials and the RHSs are posynomials. Hence, \eqref{eq: SISO_opt_3rd} is in the form of GP, which can be transformed to a convex optimization problem \cite{Boyd-Vandenberghe-04}. Now, \textbf{Algorithm} \ref{alg: SCA} can be used replacing \eqref{eq: local_conv_appr} in step 3 by \eqref{eq: SISO_opt_3rd}. Because the monomial approximation is never above the approximated summation \eqref{eq: sumgemon}, the same arguments about the convergence used in Sec.\ \ref{sec: SCAVC} can be used here. Hence, SCA with approximation \eqref{eq: SISO_opt_3rd} is guaranteed to monotonically converge to a local optimum.

\section{Numerical Results} \label{sec: results}

In this section, we will show the results obtained by the simulations to evaluate the performance of the proposed algorithms. The following abbreviations for the algorithms are used: SCAVC stands for successive convex approximation via variable change presented in Section \ref{sec: SCAVC} and SCAGP denotes successive convex approximation via geometric programming presented in Section \ref{sec: SCAGP}. The stopping criterion of \textbf{Algorithm} \ref{alg: block coordinate descent} and \textbf{Algorithm} \ref{alg: SCA} is that the change of the objective function is less than or equal to a small specific value between two consecutive iterations. In simulations, we used 0.05 for \textbf{Algorithm} \ref{alg: block coordinate descent} and 0.01 for \textbf{Algorithm} \ref{alg: SCA}.

OES stands for the best possible orthogonal allocation obtained by performing exhaustive search over all possible combinations and ZFSCMMSE denotes spatial ZF concatenated with FD-SC-MMSE. The power allocation for both OES and ZFSCMMSE is simplified to a single user loading \cite{Karjalainen-Codreanu-Tolli-Juntti-Matsumoto-11}. EP denotes the single carrier transmission without precoding, i.e., equal power is used for all users across the frequency band, where the power level satisfying the convergence constraints is found by using bisection algorithm.

The results are obtained with the following parameters: $N_F=8$, QPSK ($N_Q=2$) and 16QAM ($N_Q=4$) with Gray mapping, and systematic repeat accumulate (RA) code \cite{Divsalar-Jin-McEliece-98} with a code rate 1/3 and 8 internal iterations. The signal-to-noise ratio per receiver antenna averaged over frequency bins is defined by SNR$=\text{tr}\{\mathbf{P}\}/(N_RN_F\sigma^2)$. We consider two different channel conditions, namely, a static 5-path channel where path gains are generated randomly, and a quasi-static Rayleigh fading 5-path average equal gain channel.

\begin{figure}[tbp!]
\centering
\includegraphics[angle=-90, width=\columnwidth]{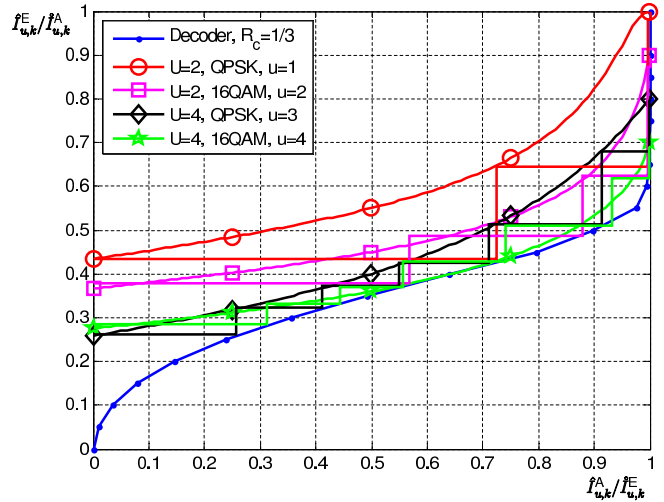}
\caption{Verification EXIT chart in static channel for SCAGP with $N_F=8$, $K=5$, $N_R=U$, $\mathring{I}_u^{\text{E,target}}=0.9999$, $\forall u$, $(\hat{I}_1^{\text{E,target}},\hat{I}_2^{\text{E,target}},\hat{I}_3^{\text{E,target}},\hat{I}_4^{\text{E,target}})
    =(0.9999,0.9,0.8,0.7)$ and $(\epsilon_1,\epsilon_2,\epsilon_3,\epsilon_4)=(0.2,0.1,0.05,0.01)$. When $U=2$, parameters of users 1 and 2 are used.}
\label{Fig: Verification}
\end{figure}

For verifying the accuracy of the method, EXIT simulations were carried out in a static channel and the trajectories were obtained through chain simulations with a random interleaver of size 240000 bits. The EXIT curve of the decoder is obtained by using 200 blocks for each a priori value with the size of a block being 6000 bits. EXIT curves of the equalizer with SCAGP and the decoder as well as the trajectories for two and four users with QPSK and 16QAM are depicted in Fig.\ \ref{Fig: Verification}. When $U=2$ and QPSK is used, the gap between the EXIT curves satisfies the preset condition and the convergence points are very close to the preset values. Furthermore, trajectory matches closely to the EXIT curves which indicates that the algorithm works properly. When the modulation order is increased to 16QAM there exists slight discrepancy between the EXIT curves and the trajectory. This happens due to the inequality \eqref{eq: upper_bound_proof}. Hence, due to the conservativeness of the convergence constraint in the case of 16QAM, the real chain simulation provides larger MI than the approximated EXIT curves and therefore, the actual trajectory reaches the convergence point. Therefore, due to the lower bound nature of convergence constraint in \eqref{eq: SIMO_opt_original} the convergence is guaranteed also with 16QAM.

To get further insight for the tradeoff between $\epsilon$ and the required SNR to satisfy the constraints we ran all the algorithms in a static channel with various $\epsilon$ and checked the SNR and the number of iterations required to achieve the target point. The results are shown in Table \ref{Tab: gap_comparison}. It can be seen that decreasing $\epsilon$ from 0.2 to 0.1 requires only one or two more iterations and the required SNR can be decreased roughly 1 dB depending on the algorithm used. The required SNR can be further reduced about 0.5 dB by decreasing $\epsilon$ to 0.01 while the number of iterations is approximately tripled.

\begin{table*}
%\vspace{-0.3cm}
\caption[]{Required SNR and number of iterations with various $\epsilon$ for algorithm. The elements in the table are in the form of SNR(dB) / iterations for user 1 / iterations for user 2. $U=2$, $N_R=2$, $N_Q=2$, $K=11$, $\hat{I}_u^{\text{A,target}}=0.9999$, $\forall u$, $\hat{I}_1^{\text{E,target}}=0.7$, $\hat{I}_1^{\text{E,target}}=0.9$.}\label{Tab: gap_comparison}\vspace*{-0.25cm}
\begin{center} \footnotesize{ %\small{ \footnotesize{
\begin{tabular} {| l | l | l | l | l | l |} %{|r|rlr|rlr|}
\hline
$\epsilon_1=\epsilon_2$ & OES & SCAGP & SCAVC & ZFSCMMSE & EP \\
\hline
0.01 & 4.56 / 23 / 16 & 4.53 / 19 / 16 & 4.54 / 17 / 17 & 6.56 / 11 / 10 & 12.79 / 2 / 2 \\ %\cline{1-3}
\hline
0.1 & 5.29 / 6 / 6 & 5.12 / 6 / 5 & 5.13 / 6 / 5 & 7.08 / 5 / 4 & 12.79 / 2 / 2     \\ %\cline{1-3}
\hline
0.2 & 6.89 / 4 / 4 & 6.28 / 4 / 4 & 6.30 / 4 / 4 & 7.96 / 3 / 4 & 12.79 / 2 / 2 \\
\hline
\end{tabular}
}
\end{center}%\vspace*{-0.6cm}
\end{table*}

\begin{figure}[tbp!]
\centering
\includegraphics[width=\columnwidth]
{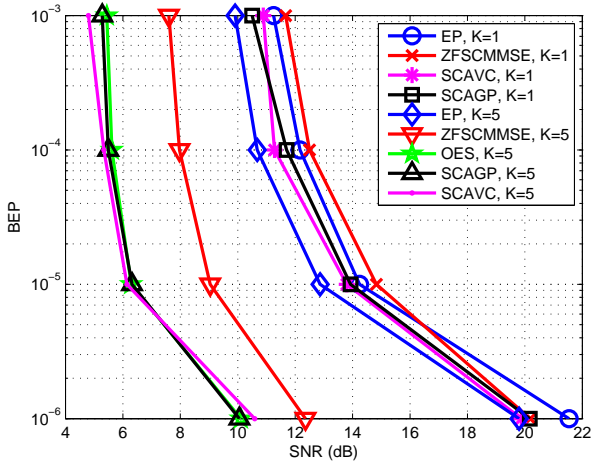}
\caption{The \textit{a posteriori} BEP comparison. $U=2$, $N_F=8$, $N_R=2$, targets = $[10^{-3}, 10^{-4}, 10^{-5}, 10^{-6}]$, $\epsilon_u=0.1$, $\forall u$.}
\label{fig: BEP_for_2_users}
\end{figure}

For QPSK, MI target can be converted to bit error probability (BEP) by using the equation \cite{tenBrink-01}
\begin{equation}
P_b\approx\frac{1}{2}\text{erfc}\Bigg(\frac{\sqrt{\text{J}_{2}^{-1}(\hat{I}_1^{\text{A,target}})
+\text{J}_2^{-1}(\hat{I}_1^{\text{E,target}})
}}{2\sqrt{2}}\Bigg).
\end{equation}
In Fig.~\ref{fig: BEP_for_2_users}, four different BEP target values were considered for $u=1,2$, namely $10^{-3}$, $10^{-4}$, $10^{-5}$, $10^{-6}$ corresponding to the MI targets $(\mathring{I}_u^{\text{E,target}},\hat{I}_u^{\text{E,target}})=(0.99, 0.6185)$, $(\mathring{I}_u^{\text{E,target}},\hat{I}_u^{\text{E,target}})=(0.9987, 0.673)$, $(\mathring{I}_u^{\text{E,target}},\hat{I}_u^{\text{E,target}})=(0.9998, 0.7892)$,
$(\mathring{I}_u^{\text{E,target}},\hat{I}_u^{\text{E,target}})=(0.9998, 0.9819)$, respectively. $K=1$ denotes the case where only one of the convergence constraints for each user is taken into account. More specifically, it means that $\hat{I}_u^{\text{A,target}}=0$, and $\hat{I}_{u,k}^{\text{E}}=\hat{I}_u^{\text{E,target}}$, $u=1,2$, $k=K$. The feedback from the decoder is not taken into account and hence, it corresponds to the linear equalizer. It can be seen that OES, SCAGP and SCAVC achieve the best result when $K=5$. ZFSCMMSE with $K=5$ is 1.77 dB - 2.9 dB worse in terms of SNR, depending on the BEP target and the algorithm used.

It is worth noticing that the solution obtained by SCAGP and SCAVC in this particular case is very close to the orthogonal solution (OES). This is due to the fact that when $\bar{\Delta}_l=0$, $\forall l=1,2,\dots,U$ in \eqref{eq: interference covariance} all the interference is canceled and the optimal receiver is the filter matched to the channel. In this case, the optimal allocation strategy to maximize \eqref{eq: zeta} is to allocate power on the strongest bin. However, this would not necessarily satisfy the constraint in \eqref{eq: SIMO_opt_original} if $\bar{\Delta}_l=1$, $\forall l=1,2,\dots,U$. Thus, the power has to be distributed to several bins which results in higher power consumption. Hence, if the tightest constraint, i.e., $\bar{\Delta}_l=1$, $\forall l=1,2,\dots,U$, can be satisfied using only one frequency bin, it is indeed the best solution. This is the case when the interference level is low, as it is in the case presented in Fig.\ \ref{fig: BEP_for_2_users}.
When the number of users increases, so does the interference and the orthogonal solution may not be feasible. This can be seen by writing the SINR constraint for OES as
\begin{equation} \label{eq: OESSINR constraint}
\frac{1}{N_F}\sum_{m\in\mathcal{N}_F^u}\frac{P_{u,m}||{\boldsymbol\gamma}_{u,m}||^2}
{P_{u,m}
||{\boldsymbol\gamma}_{u,m}||^2\bar{\Delta}_{k}+\sigma^2}\ge \xi_{u,k},
\end{equation}
where $\mathcal{N}_F^u$ is the set of frequency bins allocated to user $u$ and $\mathcal{N}_F^l\cap\mathcal{N}_F^u=\varnothing$, $\forall l\ne u$, $\bigcup_{u=1}^{U}\mathcal{N}_F^u=\mathcal{N}_F$.
Now, \eqref{eq: OESSINR constraint} can be written in the form of
\begin{align} \label{eq: minbinconstr}
&\sum_{m\in\mathcal{N}_F^u}\frac{1}
{P_{u,m}
||{\boldsymbol\gamma}_{u,m}||^2\bar{\Delta}_{k}+\sigma^2}
\le\frac{N_F^u-\xi_{u,k}N_F\bar{\Delta}_{k}}{\sigma^2},
\end{align}
where $N_F^u$ is the cardinality of the set $\mathcal{N}_F^u$.
From the nonnegativity of the right hand side (RHS) of Eq.\ \eqref{eq: minbinconstr} we get a necessary constraint for the minimum number of the frequency bins that has to be allocated to user $u$ as
\begin{equation} \label{eq: necessary constraint}
N_F^u\ge\xi_{u,k}N_F\bar{\Delta}_{k}, \;\;\;\forall k=1,2,\dots,K.
\end{equation}
As it was seen in Section \ref{sec: CCP}, $\xi_{u,k}$ and $\bar{\Delta}_{k}$ depend on the channel code used. Thus, we can conclude that the feasibility of OES algorithm can be controlled by varying the channel code. The following results are presented for 16QAM with $R_c=1/3$ only where the OES algorithm is not feasible due to \eqref{eq: necessary constraint}.

\begin{figure}[tbp!]
\centering
\includegraphics[angle=-90,width=\columnwidth]
{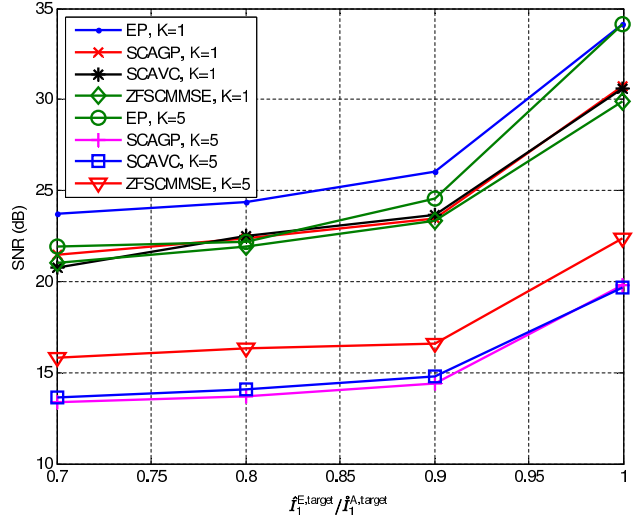}
\caption{SNR using the corresponding MI target for user 1. $U=2$, $N_F=8$, $N_R=2$, $N_Q=4$, $\hat{I}_2^{\text{E,target}}=0.8$, $\mathring{I}_u^{\text{E,target}}=0.9999$, $u=1,2$, $\epsilon_u=0.1$, $u=1,2$, $N_L=5$.}
\label{fig: SNR_for_2_users}
\end{figure}

Fig.\ \ref{fig: SNR_for_2_users} shows the minimum SNR required to achieve the corresponding MI target for user 1 for each of the proposed algorithms in the case of $U=2$.
It is shown that precoding with $K=1$ yields 5 - 8 dB worse results in terms of power consumption than the best solution with $K=5$. ZFSCMMSE with $K=1$ gives roughly the same results than SCAVC and SCAGP with $K=1$ due to the high SNR regime. However, when the precoding is performed with $K=5$, SCAVC and SCAGP achieves 2-3 dB gain compared to ZFSCMMSE.
EP with $K=5$ performs close to SCAVC, SCAGP and ZFSCMMSE with $K=1$ when the target is low. When the target is $\hat{I}_1^{\text{E,target}}=0.9999$, the EP algorithms with $K=1$ and $K=5$ are approximately equal and 3-4 dB worse than precoding with $K=1$. This is due to the fact that the scenario is interference limited, i.e., when the power is increased the interference is also increased because all the users transmit with equal power using the entire bandwidth.
As expected, EP with $K=1$ requires the highest SNR among all the algorithms used.

\begin{figure}[tbp!]
\centering
\includegraphics[angle=-90,width=\columnwidth]
{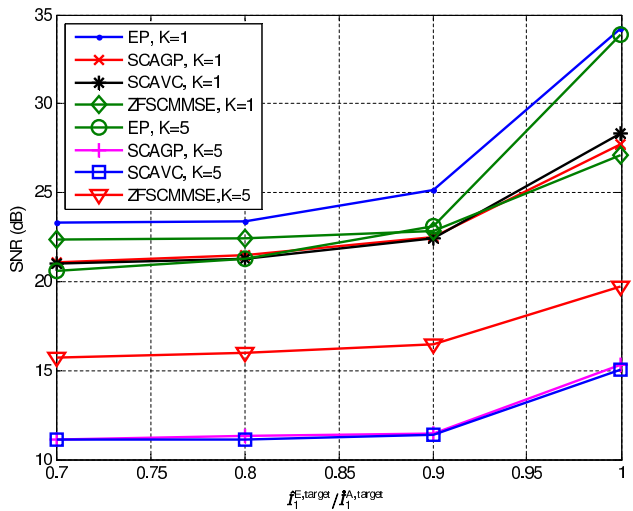}
\caption{SNR using the corresponding MI target for user 1. $U=4$, $N_F=8$, $N_R=4$, $N_Q=4$,  $\hat{I}_u^{\text{E,target}}=0.8$, $u=2,3,4$, $\mathring{I}_u^{\text{E,target}}=0.9999$, $\forall u$, $\epsilon_u=0.1$, $\forall u$, $N_L=5$.}
\label{fig: SNR_for_4_users}
\end{figure}

Fig.\ \ref{fig: SNR_for_4_users} shows the minimum SNR required to achieve the corresponding MI target for user 1 for each of the proposed algorithms in the case of $U=4$. The results are similar to the case of $U=2$: ZFSCMMSE with $K=1$ requires more power than SCAGP and SCAVC with $K=1$ when the MI target is low. However, when MI target increases ZFSCMMSE performs roughly equal to SCAGP and SCAVC. EP with $K=5$ requires smaller SNR than ZFSCMMSE when the MI target is low. The linear receivers SCAGP and SCAVC with $K=1$ are 10-13 dB away from nonlinear receivers, depending on the target MI.

As it was seen in Section \ref{sec: Algorithms} both SCAGP and SCAVC are to be solved via series of convex problems. For solving a convex problem, there exist many efficient tools \cite{Boyd-Vandenberghe-04}. Hence, the complexity analysis boils down to the comparison of how many times the optimization problem needs to be solved for each of the algorithms to achieve the convergence according to criteria described in the beginning of this section. The number of times that \textbf{Algorithm} \ref{alg: block coordinate descent} needs to be performed varies typically between 1 - 8 depending on the simulation setup. The more users, the more iterations is needed. The number of times that \textbf{Algorithm} \ref{alg: SCA} needs to be performed in \textbf{Algorithm} \ref{alg: block coordinate descent} varies between 3 - 13.

\begin{figure}[tbp!]
\centering
\includegraphics[width=\columnwidth]
{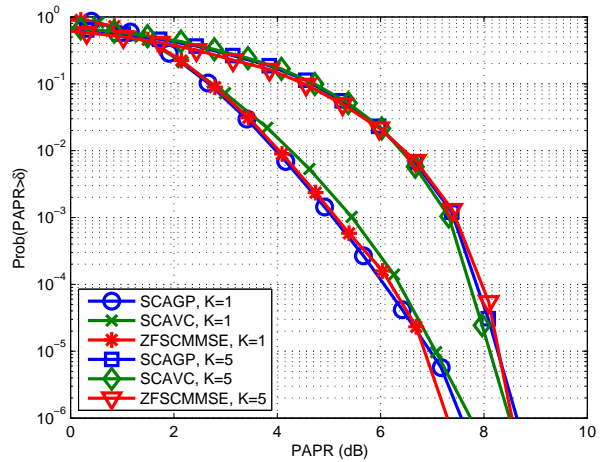}
\caption{CCDF for user 1. $U=4$, $N_F=8$, $N_R=4$, $N_Q=4$,  $\hat{I}_1^{\text{E,target}}=0.9999$, $\hat{I}_u^{\text{E,target}}=0.8$, $u=2,3,4$, $\mathring{I}_u^{\text{E,target}}=0.9999$, $\forall u$, $\epsilon_u=0.1$, $\forall u$, $N_L=5$.}
\label{fig: CCDF_for_4_users}
\end{figure}

The motivation of using SC-FDMA is its favorable PAPR properties. The PAPR of EP is only 1.27 dB for 16QAM due to the equal sizes of DFT and IDFT at the transmitter and receiver.
However, the PAPR is increased when power allocation is performed across the frequency band.
To demonstrate the effect of power allocation on the coverage of a cell, we measured the PAPR at the output of IFFT in the transmitter and constructed the complementary cumulative distribution functions (CCDF) $\text{Prob}(PAPR>\delta)$ for each algorithm. The results are shown in Fig.\ \ref{fig: CCDF_for_4_users}, where $\delta$ corresponds the PAPR value in horizontal axis. It can be seen that power allocation increases the PAPR significantly. Furthermore, with $K=5$ the PAPR is higher than with $K=1$ due to the fact that the allocation with $K=5$ is more orthogonal.
However, it can be seen from Fig.\ \ref{fig: SNR_for_4_users} that the required SNR is reduced.

Let us consider an example where the maximum transmission power is to be configured according to 8 dB PAPR which corresponds to $10^{-4.70}$ value in CCDF for SCAVC and $K=5$. For that same value of CCDF, the PAPR is 6.86 dB for SCAVC and $K=1$. Hence, increasing $K$ from one to five the total power gain is 13.22 dB - (8 dB - 6.86 dB) = 12.08 dB. Therefore, the coverage of $K=5$ precoded transmission is significantly larger than in the case of $K=1$.
However, SCAVC with $K=5$ requires 18.55 dB lower SNR than EP with $K=5$. Using the same 8 dB example than above the total power gain is 11.82 dB. However, this is only the worst case comparison, i.e., DFT and IDFT sizes are not necessarily equal in practise, which results in the increase of PAPR of EP algorithm \cite{Yuen-Farhang-Boroujeny-12}.
As a conclusion, even with the worst case comparison, SCAVC and SCAGP can achieve significantly larger coverage than EP with a significantly lower average power consumption.

\section{Conclusions} \label{sec: conclusions}

In this paper, we have derived the convergence constrained power allocation (CCPA) problem for iterative frequency domain multiuser SIMO detector. Furthermore, with our novel problem derivation the generalization for higher order modulations is straightforward. Moreover, we derived two successive convex approximations for finding a local solution of the problem. Numerical results indicate that significant gains in terms of average power consumption can be achieved compared to the linear receivers with and without precoding as well as to the iterative receiver without precoding. Furthermore, it was shown that the peak-to-average power ratio (PAPR) increase due to precoding is minor compared to the gain in the average power consumption. Thus, the maximum cell size is increased by the precoding.
Algorithms proposed in this work allow the full utilization of iterative receiver and its convergence properties.

%%%%%%%%%%%%%%%%%%%%%%%%%%%%%%%%%%%%%%%%%%%%%%%%%%%%%%%%%%%%%%%%%%%%%%%%%%%%%%%%%%%%%%%%%%%%%%%%%%%%%%%%%%%%%%%%%%%%%%%

\bibliographystyle{IEEEtran}
\bibliography{jour_short,conf_short,CCPA_MUSIMObib}

%%%%%%%%%%%%%%%%%%%%%%%%%%
\end{document}